\documentclass[journal]{vgtc}                

\ifpdf
  \pdfoutput=1\relax                   
  \usepackage{graphicx}                
  \DeclareGraphicsExtensions{.pdf,.png,.jpg,.jpeg} 
\else
  \usepackage{graphicx}                
  \DeclareGraphicsExtensions{.eps}     
\fi%

\graphicspath{{figures/}{pictures/}{images/}{./}} 

\usepackage{microtype}                 
\PassOptionsToPackage{warn}{textcomp}  
\usepackage{textcomp}                  
\usepackage{mathptmx}                  
\usepackage{times}                     
\usepackage{cite}                      
\usepackage{tabu}                      
\usepackage{booktabs}                  
\usepackage{soul} 
\usepackage{array}
\newcolumntype{M}[1]{>{\centering\arraybackslash}m{#1}}

\usepackage{xcolor}

\newcommand{\rev}[1]{{\color{black}{#1}}}

\usepackage{subcaption}

\usepackage{balance}

\onlineid{1157}

\vgtccategory{Research}
\vgtcpapertype{Evaluation}

\title%
{      
       A Comparison of Radial and Linear Charts for \\Visualizing Daily Patterns
        }
        
\author{
Manuela Waldner, \textit{Member, IEEE CS}, Alexandra Diehl, Denis Gra\v{c}anin, \textit{Senior Member, IEEE},\\Rainer Splechtna, \textit{Member, IEEE CS}, Claudio Delrieux, and Kre\v{s}imir Matkovi\'{c}, \textit{Member, IEEE CS}
}

\authorfooter{
\item
Manuela Waldner is with TU Wien. E-mail: waldner@cg.tuwien.ac.at.
\item
Alexandra Diehl is with University of Zurich. E-mail: diehl@ifi.uzh.ch.
\item
Denis Gra\v{c}anin is with Virginia Tech. E-mail: gracanin@vt.edu.
\item
Rainer Splechtna is with VRVis Research Center. E-mail: Splechtna@VRVis.at.
\item
Claudio Delrieux is with Electric and Computer Eng. Dept., Universidad Nacional del SUR and CONICET. E-mail: cad@uns.edu.ar.
\item
Kre\v{s}imir Matkovi\'{c} is with VRVis Research Center. E-mail: Matkovic@VRVis.at.
 }

\shortauthortitle{Waldner et al.: Comparison of Radial and Linear charts for Visualizing Daily Patterns}

\abstract{
Radial charts are generally considered less effective \rev{than} linear charts. 
Perhaps the only exception is \rev{in} visualizing periodical time-dependent data, which is believed to be naturally supported by the radial layout. 
It has been demonstrated that the drawbacks of radial charts outweigh the benefits of this natural mapping.
\rev{Visualization of daily patterns, as a special case,} has not been \rev{systematically evaluated} using radial charts.
In contrast to yearly or weekly recurrent trends, the analysis of daily patterns on a radial chart may benefit from our trained skill on reading radial clocks that are ubiquitous in our culture. 
In a crowd-sourced experiment with 92 non-expert users, we evaluated the accuracy, efficiency, and subjective ratings of radial and linear charts for visualizing daily traffic accident patterns.
We systematically compared juxtaposed 12-hours variants and single 24-hours variants for both layouts in four low-level tasks and one high-level interpretation task.
Our results show that over all tasks, the most elementary 24-hours linear bar chart is most accurate and efficient and is also preferred by the users.
This provides strong evidence for the use of linear layouts -- even for visualizing periodical daily patterns.
} 


\CCScatlist{ 
 \CCScat{K.6.1}{Management of Computing and Information Systems}%
{Project and People Management}{Life Cycle};
 \CCScat{K.7.m}{The Computing Profession}{Miscellaneous}{Ethics}
}

\begin{document}


\firstsection{Introduction}

\maketitle

Rose charts, also called coxcomb charts or polar-area charts, became popular in the 19$^{\mathrm{th}}$ century when Florence Nightingale visualized mortality in the Crimean War from April 1854 to March 1855~\cite{cohen1984florence,brasseur2005florence}.
The charts show the number of soldier deaths for each month split into two rose charts. 
Since then, rose charts or similar radial charts have been frequently used to depict periodical temporal patterns, such as frequencies of daily news reports \cite{sheidin2017visualizing}, 24-hours network traffic~\cite{erbacher2002intrusion,kintzel2011monitoring}, distribution of phone calls over the day~\cite{shen2008mobivis}, daily traffic information~\cite{huang2016trajgraph}, climate effects over a year \cite{seebacher2018visual}, or monthly sales~\cite{ren2019charticulator}.
It is often argued that mapping periodical behaviors to a radial chart is an intuitive and appropriate representation of the underlying data~\cite{burch2014benefits,kintzel2011monitoring,brasseur2005florence}. 

However, there is empirical evidence that radial charts are inferior for many analytic tasks as compared to linear ones~\cite{cleveland1984graphical,heer2010crowdsourcing,simkin1987information,goldberg2011eye,burch2014benefits,blascheck2018glanceable}.
This has even been shown for radial charts encoding periodical time-dependent data, where linear charts outperformed or performed equally well as radial ones when conducting low-level analytic tasks~\cite{adnan2016investigating,brehmer2017visualizing}.
These studies, however, compared weekly, monthly, or yearly periodical patterns.
\rev{To the best of our knowledge, Fuchs~et~al.~\cite{fuchs2013evaluation} presented the only user} study comparing different glyph layouts for monitoring tasks of \emph{daily} patterns.
\rev{They} found that radial glyphs are more effective for reading values at specified time points.
In particular, their users reported to like the clock metaphor of the color-coded 24-hours clock glyph.
\rev{However, such a glyph does not utilize the layout of an analog clock. 
For future work, some of their users therefore suggested to visualize twelve hours per glyph to create a more intuitive mapping. }



Hours of a day are \rev{a} special \rev{domain,} as observers are trained to read the hours from radial clocks from an early age\rev{~\cite{friedman1989children, boulton1997analysis}}.
It is therefore possible that familiarity with the radial clock design facilitates a more efficient cognition of daily patterns in the underlying data when visualizing the data with a radial \rev{clock-like} layout.
\rev{This is particularly relevant for non-expert audiences, where the goal is to communicate information of general interest, such as daily traffic information or social media presence over the day, effectively, but also in an engaging manner~\cite{lankow2012infographics}}.  

\rev{We present the first controlled user study investigating whether imitating a radial analog clock layout can facilitate the interpretation of \emph{daily} patterns.}
Our main hypothesis was that \rev{visualizations of daily patterns using a clock-like radial layout can be interpreted more efficiently than using alternative radial or linear charts by untrained users due to the familiar layout}.
\rev{We conducted} a crowd-sourced experiment where we compared the users' performance with two layouts (linear and radial) with two different encoding cardinalities (2x12 hours and 1x24 hours).
Thereby, the two juxtaposed radial rose charts use the same encoding principle as a radial clock, where the twelve hours of a.m.~and p.m.~are arranged in one circle each.
The single 24-hours radial variant, where noon is located at the bottom and midnight at the top, can also be observed in \rev{many} scientific works (e.g.,~\cite{sheidin2017visualizing,huang2016trajgraph, fischer2012clockmap,masoodian2013time,seebacher2018visual,shen2008mobivis}) and in-the-wild (e.g.,~\cite{creativeRoutines}), but does not correspond to the familiar 12-hours clock representation.
\rev{We contribute evidence that a clock-like radial layout does not have a positive impact on understanding daily patterns for untrained users.
We also provide new findings about the effect of chart separation on users' task performance for both linear and radial charts.
Finally, we performed an exploratory analysis of error cases and report observations that can explain why untrained users often misinterpret rose charts. }

\section{Background}
\label{sec:background}


Radial charts are often considered more aesthetic and space-efficient~\cite{heer2010tour,kintzel2011monitoring, burch2014benefits}, as well as more natural and therefore more memorable~\cite{borkin2013makes} \rev{than} linear charts. 
They are commonly used to visualize hierarchical data~\cite{stasko2000focus+}, multivariate data~\cite{hinum2005gravi++,zhao2011kronominer}, or set relations~\cite{alsallakh2013radial}. 
Draper~et~al.~\cite{draper2009survey} provide a survey of radial charts up to 2009. 
Of our special interest is their presumable intuitiveness for visualizing time-dependent periodical data~\cite{burch2014benefits,kintzel2011monitoring,brasseur2005florence}. 

Rose charts~\cite{cohen1984florence,brasseur2005florence} can be seen as radial histograms of time-dependent data, where each bar represents the frequency for a discrete temporal bin, such as one month or one hour. 
Circular silhouette graphs encode frequencies as continuous curve following a yearly circle~\cite{harris2000information}. 
Brehmer~et~al.~\cite{brehmer2017visualizing} visualize value ranges as bars, both in a linear and a radial layout. 
Alternatively to bars, values can be represented by lines in polar coordinates~\cite{erbacher2002intrusion}.
The peaks of these star glyphs can additionally be connected by lines~\cite{fuchs2014influence}. 
Finally, time-dependent values can be encoded by color resulting in clock glyphs~\cite{fuchs2013evaluation} or ``Kaleidomaps''~\cite{bale2007kaleidomaps}. 

An alternative to juxtaposition of multiple radial graphs to show longer time sequences, like the two years in Nightingale's rose charts~\cite{cohen1984florence,brasseur2005florence}, is to arrange longer time sequences as concentric rings~\cite{chuah1998dynamic,drocourt2011temporal} or in a spiral~\cite{dragicevic2002spiraclock}.
This principle can also be used for visualization of longer time series data, where each year, for instance, is visualized in 360 degrees~\cite{weber2001visualizing,tominski2008enhanced}. 

\rev{Several studies indicated that radial charts are less effective than linear charts for exploratory tasks, relative judgments, and readability.} 
For instance, it has been shown that relative judgments are more difficult to perform using pie charts than bar charts~\rev{\cite{cleveland1984graphical,simkin1987information,heer2010crowdsourcing,few2007save} because area and angle judgments are less accurate than position or length judgments~\cite{cleveland1984graphical}. 
In addition, non-salient angles ({\em i.e.}, not 0$^{\circ}$, 90$^{\circ}$, etc.) are hard to interpret in pie charts \cite{simkin1987information,few2007save}. }
For small-scale charts on smart watches, Blascheck~et~al.~\cite{blascheck2018glanceable} showed that classic linear bar charts are \rev{also} easier to read than donut charts or radial bar charts, where bars are represented as arcs. 
Goldberg and Helfman~\cite{goldberg2011eye} compared different variations of linear and radial fixed-sized charts in an eye tracking study.
They could show that reading values is clearly more difficult in radial charts \rev{because mapping data points to values is more inefficient compared to linear charts.}
\rev{Through a large-scale online study, Diehl et al.~\cite{diehl2010uncovering} showed that memorizing specified locations is also more difficult for radial visualizations than Cartesian visualizations, but circle sectors are easier to memorize than rings.}
\rev{On the other hand side, Spence~\cite{spence2005no} showed that radial charts perform as well as linear charts for comparing proportions, and that they provide better visual cues for part-whole estimations and natural anchors (0$^{\circ}$, 90$^{\circ}$ etc.). }These studies did not focus on time-dependent data.
 








Several studies compared linear visualizations for time series data.
Javed~et~al.~\cite{javed2010graphical} found that separating multiple time series into small multiples is more effective than superimposing them. 
Gogolou~et~al.~\cite{gogolou2018comparing} compared three linear time series visualization techniques (line charts, horizon charts, and color fields) and  concluded that the visual encoding has a lot of influence on the perception of similarity between two time series.
Albers~et~al.~\cite{albers2014task} compared different linear visualization techniques for different tasks analyzing time series data. 
They found that a composite graph, overlaying a line graph with a bar chart representing the averages of the bar's region, outperformed other time series representations. 

Of most relevance for the encoding of daily patterns are studies comparing linear and radial layouts for periodical time series data.
Adnan~et~al.~\cite{adnan2016investigating} compared different linear and radial layouts for visualizing time series data across several weeks. For all tasks, the linear charts outperformed the radial charts. 
Brehmer~et~al.~\cite{brehmer2017visualizing} systematically compared linear and radial \rev{bar} charts for visualizing weekly, monthly, and yearly patterns on mobile phones.
They found that linear charts were generally more efficient, while the accuracy was comparable to radial charts.
Our study differs as we look into the special case of daily patterns that may benefit from the visualization in a clock-like chart. 

To the best of our knowledge, Fuchs~et~al.~\cite{fuchs2013evaluation} presented the only study comparing different glyph designs, including linear and radial variants, for \emph{daily} patterns in different data granularities from 24 to 96.
They \rev{systematically} compared \rev{four} different glyphs\rev{: radial star and color-coded clock glyphs and linear line and colored stripe glyphs.} 
They found that linear glyphs were most accurate and efficient for value comparison, but the \rev{radial} glyphs were more accurate and efficient, respectively, for time comparisons.
Informal feedback \rev{by their student participants} suggests that the clock glyph was appreciated for its obvious clock metaphor.
This inspired us to investigate this aspect in more detail by systematically varying both, the layout \rev{(linear and radial)} and the encoding granularity, to test the effect of the clock design on the task performance with large-scale visualizations rather than glyphs.
\rev{As we were interested in the appropriateness of clock-like radial charts for showing daily patterns for a broad audience, a difference of our study is that we performed a crowd-sourced study with potentially untrained users. 
The population registered at crowd-sourcing platforms is expected to differ from student subjects in multiple aspects, such as a wider age range and higher openness (imagination, curiosity) \cite{kosara2010mechanical}. }

\section{Visualization of Daily Patterns}
\label{sec:encodings}

The variety of techniques to visualize periodical time-dependent patterns in Section~\ref{sec:background} shows that there is a large possible design space.
This design space includes linear charts, such as line charts, bar charts, or colorfields, as well as radial charts, such as star glyphs, rose charts, or color-based clock glyph\rev{s} (Table~\ref{tab:designSpace}).
In this design space, we only consider radial charts where the time is encoded in angle and the value in radius or color. 
Radial bar charts sometime also use angle to encode the value (e.g.,~\cite{blascheck2018glanceable}), but we do not consider this case here.

\begin{table}[]
\centering
\caption{Design space of linear and radial visualizations for periodical time-dependent patterns, organized by encoding (rows) and layout (columns). \rev{In our study, we investigated bar encodings (second row).} }  
\label{tab:designSpace} 
\begin{tabular}{|m{1cm}|M{2.5cm} M{2.5cm}|}
\hline
\textbf{} & \textbf{linear} & \textbf{radial}   \\ \hline
 \textbf{line} & \includegraphics[scale=.13]{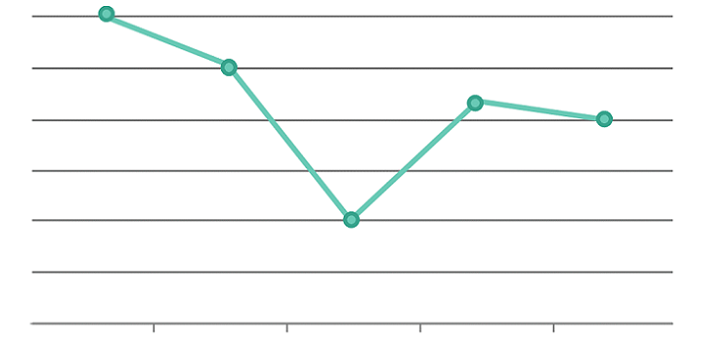} & \includegraphics[scale=.30]{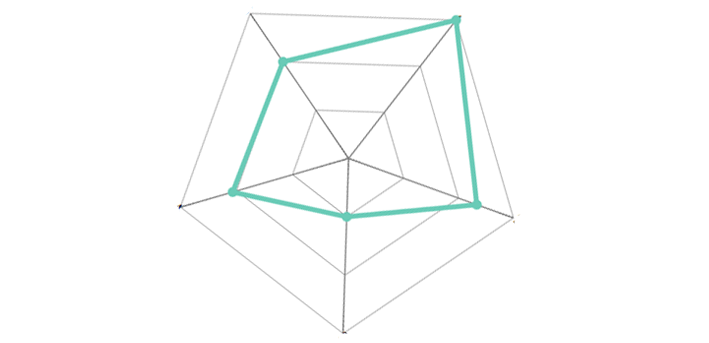}  \\
 \textbf{bar} & 
 \includegraphics[scale=.13]{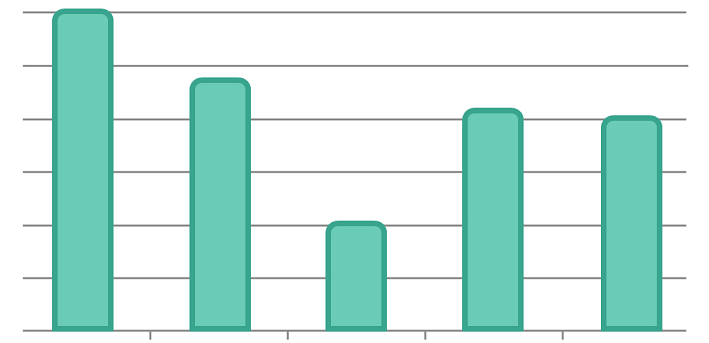} & \includegraphics[scale=.30]{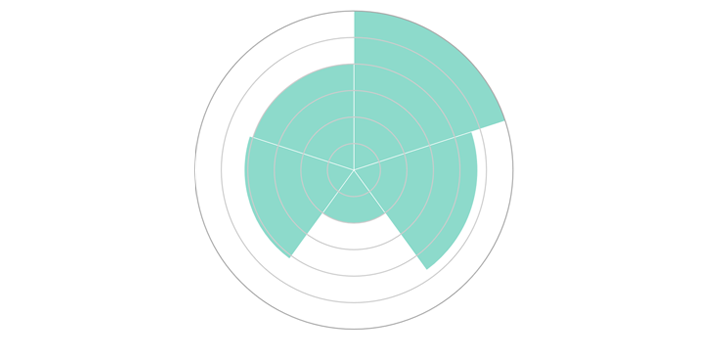} \\
 \textbf{color} & \includegraphics[scale=.25]{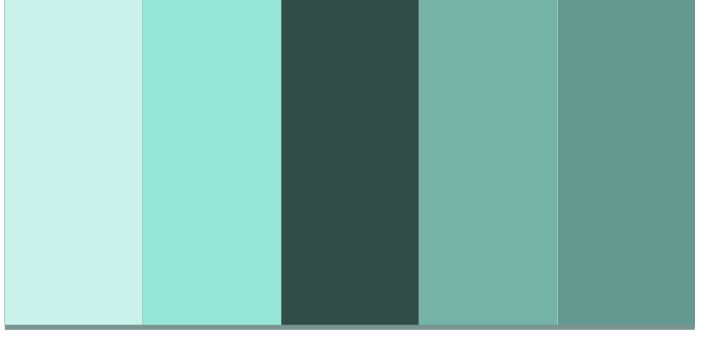} & \includegraphics[scale=.30]{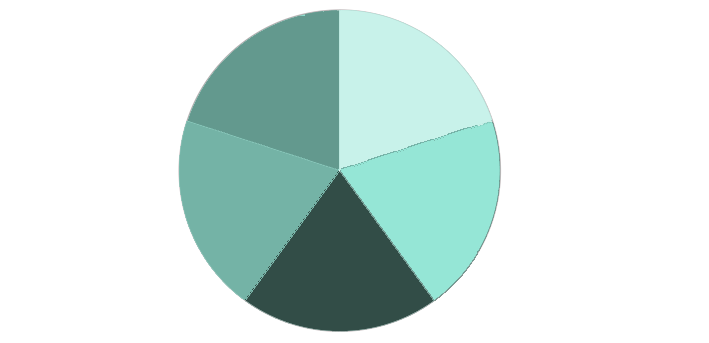} \\
 \hline
\end{tabular}
\end{table}

Our goal \rev{was} to visualize aggregate values across 24 hours, such as the accumulated energy consumption per hour or the frequency of traffic accidents for each hour a day.
This corresponds to a \rev{periodical} temporal histogram with 24 bins. 
We excluded color as it is expected to be inferior \rev{to position encodings \cite{albers2014task}.}
Goldberg and Helfman~\cite{goldberg2011eye} showed that across linear and radial charts, data points could be found more efficiently using bar charts than line charts.
Also, as we are dealing with binned data, interpolating data points with line connections could lead to misinterpretations.
According to Borkin~et~al.~\cite{borkin2013makes}, bars -- including linear bar charts, stacked bar charts, and circular bar charts -- are also the most commonly used encoding in news media and government reports.
We therefore chose a classic bar histogram for visualizing our data. 
In polar coordinates, this corresponds to a rose chart. 


This encoding choice leaves us with four simple visualizations for 24-hours time series data: two juxtaposed 12-hours radial charts (\textbf{12r}), a single 24-hours radial chart (\textbf{24r}), two juxtaposed 12-hours linear bar charts (\textbf{12l}), and a single 24-hours linear bar chart (\textbf{24l}).
The 12-hours radial chart (12r) adopts the ubiquitous clock layout with twelve on the top and six at the bottom.
At the same time, it suffers from the known disadvantages of radial charts, such as the difficulty to decode values accurately and efficiently. 
\rev{In addition, to encode the 24 hours of the day, the data has to be split into two separate charts. This means that, in contrast to the 24-hours radial chart, the data cannot be read without switching charts.}
This is also true for the two 12-hours linear bar charts (12l).
However, in this case, the right-most bar of the first 12-hours chart marks a clear end of the a.m.~section. 
Continuing to the beginning of the p.m.~section is supported by the natural reading order.
On the other hand, the separation into a.m.~and p.m.~may also be advantageous as the visualized data is separated into smaller chunks with two identical value axes. 
The 24-hours radial chart (24r) is the only visualization that is truly continuous so that also temporal patterns across midnight are not separated.



\begin{figure*}
	\centering
	\begin{subfigure}{0.6\textwidth}
		\includegraphics[width=4.1in]{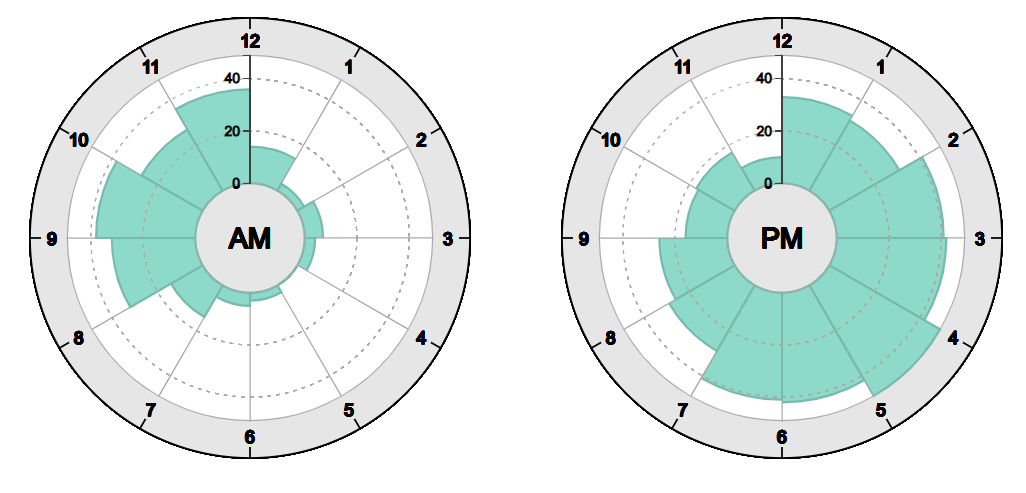}
		\caption{12-hours radial (12r)} 
		\label{fig:12r}
	\end{subfigure}
	\begin{subfigure}{0.3\textwidth} 
		\includegraphics[width=2.0in]{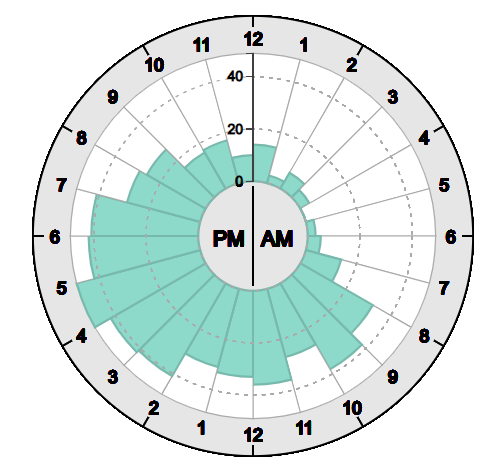}
		\caption{24-hours radial (24r)} 
		\label{fig:24r}
	\end{subfigure}
	\caption{2x12-hours and 24-hours radial charts showing the number of traffic accidents in Staten Island, New York, in March. } 
	\label{fig:radial}
\end{figure*}

\section{Experiment}
\label{sec:experiment}

\rev{The goal of this experiment was to verify our hypothesis that daily patterns can be interpreted most efficiently when visualized with a clock-like radial layout.}
\rev{We} also hypothesized that for a more detailed analysis, like reading or comparing values, a linear layout will be superior, as suggested by prior work~\cite{goldberg2011eye,adnan2016investigating,brehmer2017visualizing}.
Therefore, it is sensible to hypothesize that there is no benefit for a 24-hours radial chart, as it suffers from known decoding deficiencies, and at the same time does not provide the clock metaphor benefit of the 12-hours radial chart.
As our goal was to investigate how well visual encodings are understood and accepted by a general audience, we crowd-sourced our experiment using Amazon's Mechanical Turk system, commonly used to perform graphical perception studies~\cite{heer2010crowdsourcing,kosara2010mechanical,borgo2018information}. 
\rev{Also, previous studies comparing time series visualizations have been conducted remotely through Mechanical Turk \cite{albers2014task,brehmer2017visualizing}.}


\subsection{Visualizations}
\label{sec:expEncodings}

\begin{figure}[tb]
\centering
\includegraphics[width=0.8\columnwidth]{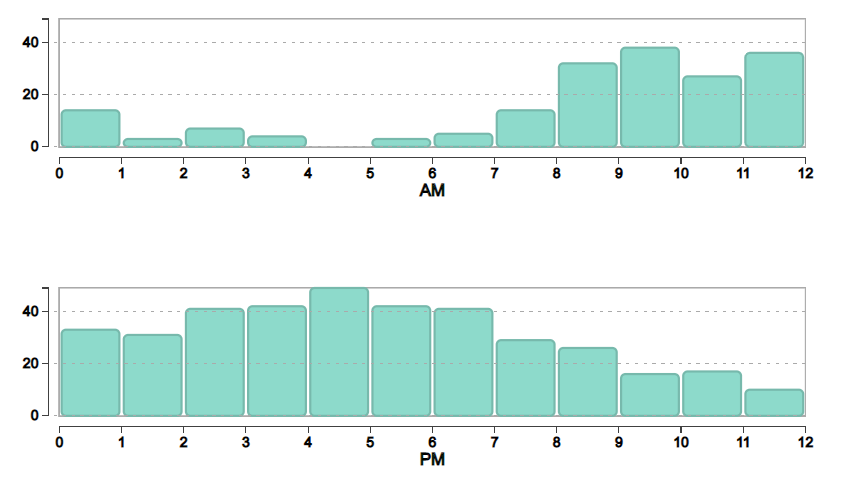}
\caption{2x12-hours linear bar charts (12l).}
\label{fig:12l}
\end{figure}

\begin{figure}[tb]
\centering
\includegraphics[width=0.8\columnwidth]{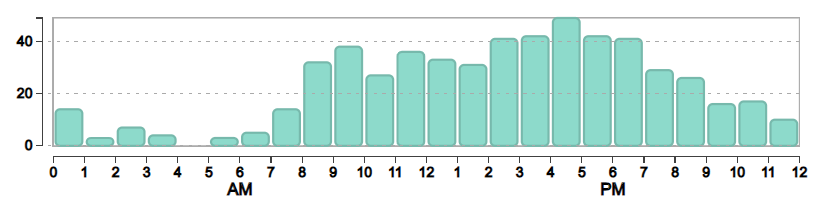}
\caption{24-hours linear bar charts (24l).}
\label{fig:24l}
\end{figure}

To test these hypotheses, we created four visualizations: two juxtaposed 12-hours radial charts (12r, Fig.~\ref{fig:12r}), a single 24-hours radial chart (24r, Fig.~\ref{fig:24r}), two juxtaposed 12-hours linear bar charts (12l, Fig.~\ref{fig:12l}), and a single 24-hours linear bar chart (24l, Fig.~\ref{fig:24l}).
These visualizations differ in two aspects: their encoding \textbf{cardinality}, which can be 12 hours or 24 hours, and \textbf{layout}, which is radial or linear. 
\rev{We recruited our study participants from North America. Therefore, we encoded time in the 12-hours clock system using AM~and PM~labels.}


As our focus was to investigate visualizations for the masses, we used a constant chart size, representative of visualizations found in news or on websites. 
Borkin~et~al.~\cite{borkin2013makes} used up to $512\times512$ pixels for their representative in-the-wild visualizations in the study. 
Similarly, we
rendered the two juxtaposed 12-hours radial charts with a radius of 200 pixels. 
We left a 60 pixel radius inner circle, where we placed the AM~and PM~labels. 
The outer ring, where the hour labels are located, has a darker color and uses additional tick marks to emphasize the clock layout. 
\rev{Like the original Nightingale rose chart}, the two 12-hours radial charts were rendered next to each other \rev{to make the best use of conventional monitor aspect ratios}. \rev{On small screens, they were rendered} on top of each other. 

The 24-hours radial chart is rendered with the same radius. 
This makes the bars narrower compared to the two 12-hours charts.
Like for the 12-hours variant, the bar value is encoded along the radial axis.
We decided to keep the radius constant to keep the value encoding constant across the conditions. 
This means that the 24-hours radial chart is more space efficient compared to the 12-hours variant. 
In the inner circle, we placed the label ``PM~$|$AM'' for the two circle halves.  

The linear bar charts (12l, 24l) were rendered with a height of 140 pixels.
In this way, the y-axis scale is the same size as the scale in the radial charts. 
The width of the bar charts is the same as the circumference of the circle, which corresponds to the medium value grid line in the radial charts.  
The two 12-hours linear bar charts were rendered with the same width as the 24-hours bar chart to keep the difference between 12-hours and 24-hours variants consistent between radial and linear layouts.

Both linear bar charts have AM~and PM~labels centered underneath the hour label ``6''. 
The 12-hours bar charts have two identical y-axes for both charts. 
In contrast to the 12-hours radial charts, we always rendered the two 12-hours bar charts on top of each other for two reasons:
First, the difference to the 24-hours linear chart is minimal when rendering the two charts next to each other. 
Second, with a vertical alignment, AM~and PM~values are placed directly on top of each other for easier comparison. 


\subsection{Data}
\label{sec:data}

As our experiment also included a high-level analysis task (Section~\ref{sec:task1}), 
we needed a generally understandable data set. 
We used New York traffic accident data from 2013 to 2016~\cite{NYC_DATA}.
We binned that data into boroughs, months, and hours of the day so that each bin gives the accumulated number of accidents per month and per hour of the day for a selected borough.

For the final experiment, we 
\rev{used Staten Island borough data from 2016}, as the number of traffic accidents and their distribution across the time of the day varied most among months \rev{(compared to other boroughs). 
These twelve months have} distinct patterns to circumvent the learning effect when performing repeated measures. 
We did not systematically vary the data as a prior experiment comparing radial and linear charts \rev{by Brehmer et al.~\cite{brehmer2017visualizing}} has already shown that the data characteristics have no influence on the users' performance, as long as the periodic nature of the visualization fits the data (e.g., sleep patterns over the week, or temperature distributions over the year).


\subsection{Tasks}
\label{sec:tasks}

We included one high-level analysis task, four low-level analytic tasks, and one subjective rating task to test our hypotheses.
The four low-level analytic tasks (\rev{tasks 2--5}) were adopted from Brehmer~et~al.~\cite{brehmer2017visualizing}. 
\rev{Some of these low-level analytical tasks, such as finding the maximum bar, do not take the temporal component into account. For these tasks, the influence of the clock-like layout therefore can be considered negligible. However, as rose charts are slightly different from the radial bar charts used in the study by Brehmer~et~al.~\cite{brehmer2017visualizing}, replicating these results helps to generalize their findings to other types of radial charts.}
Every task was preceded by a task description before showing the actual visualization. 
Below, we explain the tasks together with our outcome hypotheses and the measures we obtained to verify the hypotheses. 

\subsubsection{Task 1: Untargeted Analysis}
\label{sec:task1}

The purpose of this task was to simulate confronting with the visualization in-the-wild, such as on a news website, without any prior instructions how to read the visualization. 
\rev{Therefore, }we did not provide any visualization reading instructions. 
Before showing the visualization, we only informed the users that we would very briefly show how the number of traffic accidents in Staten Island is distributed \rev{during} the 24 hours of the day.
We asked the users to carefully inspect the visualization \rev{during} the few seconds that it will be visible, and afterwards to write down everything they observed, even if they are unsure.
We showed the \rev{non-interactive} visualization for ten seconds, 
\rev{after which it was replaced by} a text field, where users could take note of their observations. 
Using this task, our goal was to probe which high-level observations untrained users made within a short period of time. 
This includes observations of peaks, comparisons of longer periods (such as morning compared to afternoon), or minima.
Our hypothesis was that \emph{the 12-hours radial layout (12r) will lead to more observations than any other layout} (\textbf{H1}), as the familiarity with the clock layout will facilitate the identification and relation of time periods so that general trends and patterns can be perceived more efficiently. 

\subsubsection{Task 2: Locate Time}

\rev{The goal of this task was to assess how quickly and accurately users can find a specified hour of the day in the visualization. We randomly picked a one-hour period (e.g., 8--9 a.m.) and asked the user to click on the bar corresponding to the specified time interval. 
This one-hour interval was randomly selected from 22 hours of the day. 
We} explicitly excluded the periods between twelve and one to avoid confusion between 12 a.m.~and 12 p.m.
Our hypothesis was that \emph{selecting a one-hour interval on a 12-hours chart (12r, 12l) is more efficient than on a 24-hours chart} (\textbf{H2}).
The rationale for this is that separation into a.m.~and p.m.~would facilitate the search for the target time value compared to a single, continuous encoding.
In accordance with user feedback given during a study by Fuchs~et~al.~\cite{fuchs2013evaluation}\rev{, and because children are usually able to read hour intervals from analog clocks from age five to seven \cite{friedman1989children,boulton1997analysis,burny2013curriculum}}, we furthermore hypothesized that \emph{the familiar radial clock layout (12r) will facilitate the location of time intervals compared to the linear layout (12l)} (\textbf{H2}). 
We logged which bar the user selected, as well as the time between visualization onset and bar selection. 

\subsubsection{Task 3: Read Value}

The aim of this task was to read a value for a specified hour.
To separate the reading task from the interval location task, the target interval was specified by an arrow.
The target interval was selected randomly from all 24 hour intervals. 
Based on prior empirical evidence\rev{s \cite{goldberg2011eye,brehmer2017visualizing}}, our hypothesis was that \emph{reading values is more accurate and efficient using a linear layout (12l, 24l) than a radial layout.} 
Due to the separation into two bar charts with their own y-axes, we hypothesized that \emph{reading values is most accurate and efficient using the 12-hours linear layout (12l)} (\textbf{H3}).
We logged the user responses and the time from visualization display to selection of the ``continue'' button. 
To quantify the accuracy, we computed the absolute difference between the user response and the actual hour value. 

\subsubsection{Task 4: Locate Maximum}

For this task, we asked users to quickly select the maximum value from the visualization. 
In contrast to Brehmer~et~al.~\cite{brehmer2017visualizing}, we did not query the minimum value as our dataset had a lot of very low or even zero hour values so that this task was not meaningful in our case. 
Our hypothesis was that \emph{finding the maximum value is most efficient using the 24-hours linear layout (24l)} (\textbf{H4}), since in this continuous, linear arrangement, the maximum value clearly stands out. 
We logged the value of the selected bar and the time until the bar was selected.

\subsubsection{Task 5: Compare AM/PM~Interval Values}

Users were asked to decide as to whether in a one-hour time interval marked by a blue arrow, there were fewer, the same number, or more traffic accidents than in the reference hour marked by a red arrow. 
The response had to be selected from three radio buttons beneath the visualization.
The location of the blue arrow was selected randomly from all 24 hours. 
The red arrow was then shifted exactly for twelve hours \rev{forwards or backwards}. 
This means, we always compared the same time intervals in a.m.~and p.m. 
We made this decision as different intervals may confound with the layout and the cardinality. 
For instance, we can expect an interaction effect between cardinality and layout for comparing adjacent intervals depending on whether the adjacent intervals are both within a.m.~or distributed among a.m.~and p.m. 
In order not to test all these possible arrangements, we decided to stick to this one specific comparison case. 

We hypothesized that \emph{any linear layout (12l, 24l) facilitates a more effective comparison of a.m./p.m.~interval values} (\textbf{H5}). 
The 24-hours bar chart benefits from the common baseline of the linear bars, while the 12-hours bar charts ensure that common a.m./p.m.~intervals are always located on top of each other.
We logged the time between visualization display and selection of the ``continue'' button, the actual value difference, as well as the user's selected radio button. 
As a correctness measure, we computed if the selected radio button corresponded to the sign of the value differences.

\subsubsection{Task 6: Subjective Rating}

\rev{We} asked users to state their subjective assessment for the respective visualization to show traffic accident frequencies on a Likert scale from 1 (``it is not suitable at all'') to 5 (``there is no better way to show it''). 
Our hypothesis was that \emph{users rate the 24-hours radial layout (24r) lowest} (\textbf{H6}) as it has no apparent advantages for the presented tasks. 

\subsection{Design}
\label{sec:design}

We used a mixed design with the two independent variables \textbf{layout} (radial or linear) as within-subjects factor and encoding \textbf{cardinality} (12 or 24) as between-subjects factor. 
This means, every user performed all six tasks with both, a radial and a linear layout, resulting in 6 (tasks) $\times$ 2 (layouts) = 12 trials per user in total.
We chose encoding cardinality as between-subjects factor as we believe that switching the cardinality with the same layout during the study may lead to misinterpretations, while switching between linear and radial layout is sufficiently different so that users can readily adjust their decoding strategy.
The dependent variables were the number of observations (task 1), accuracy and task completion time (tasks 2-5), and subjective rating (task 6).

The six tasks were always presented in the same order as listed in Section~\ref{sec:tasks} -- first for the radial layout, then for the linear layout, or vice versa. 
The presentation order of radial and linear was balanced between the workers.
The twelve data sets described in Section~\ref{sec:data} were assigned to the twelve tasks in random order. 
With task~1 as the first task for a given layout, we could test whether users would understand the layout and get some first impression of the data without any further guidance -- as if they would be confronted with the visualization in-the-wild.
For the subsequent tasks, where completion time was a concern, they were therefore already familiar with the encoding.

\subsection{Apparatus}
\label{sec:apparatus}

Users signed up for a single \rev{Human Intelligence Task (HIT)} with a link to an external server.
We performed a pilot study with five users, \rev{a mix of visualization experts and} non-experts. 
The average study completion time was 15 minutes. 
We therefore configured the HIT to stay active for 30 minutes. 
The visualizations were created using the D3 JavaScript library. 
Users were assigned an anonymous PHP session ID, which was logged together with their responses on the external server. 
The server script sequentially assigned workers to one of the four possible configurations (cardinality 12 and radial first, etc.). 

As most of our tasks were rather simple to solve, we inspected the users' responses to decide whether to accept the HIT for the user. This is a common quality assurance procedure for crowd-sourced studies \cite{borgo2018information}.
Upon completion of the study, the server displayed a code containing the session ID and an estimated error code, which users had to copy to the Mechanical Turk response field. 
The error code was increased if the text field for task 1 was empty, the selected bar in task 2 was incorrect, the entered value or selected maximum value in task 3 and task 4, respectively, deviated more than 10\% from the actual value, and if the rather simple comparison task 5 was answered incorrectly.
When users received an error code higher than 2, we manually checked the server logs for their responses and excluded cases that did not provide any meaningful qualitative observations in task 1. 
Users received \rev{a 2 USD} compensation if they completed all 12 tasks and the quality criteria above were fulfilled. 
On average, users required 13 minutes to complete the study. 

\subsection{Participants}
\label{sec:participants}

We set 100 users as target for our HIT, which is within the common range of participants in crowd-sourced studies~\cite{borgo2018information} \rev{and was also used by the methodologically most similar prior study to ours by Brehmer et al.~\cite{brehmer2017visualizing}}.
\rev{We only allowed participants from North America. In the USA, telling and writing time is taught in maths courses from the first grade \cite{commonCore}. }

From the 100 obtained responses, we rejected eight because they did not provide meaningful responses to task 1 and 
had more than two incorrect responses in total (see Section~\ref{sec:apparatus}). 
Age ranged from 19 to 68 with a mean age of 36.2, and 41.3\% of the users were female.
We did not perform a color-blindness test, since color was not used to encode data. 
As the labels of the visualizations were larger than the font used for the introduction and task description, we also did not perform any additional visual acuity screening but rather expected anybody able to read the instructions to be a plausible reader of the visualization. 
Most users reported that they are not at all (19.6\%) to averagely (51\%) familiar with information visualization. 
Eight users stated that they are quite familiar, but no user claimed to be a visualization expert. 

\subsection{Analysis} 

For the untargeted analysis (H1), we performed a qualitative data analysis using open coding of the obtained text responses with three \rev{independent human} coders.
In the first step, we established a codebook \rev{to characterize high-level observations by the users.}
The coders \rev{iteratively adjusted the codebook and discussed conflicting codes} until they reached an inter-coder reliability of $>0.8$ according to Krippendorff's alpha. 
\rev{The final codebook contained the following codes:}
\begin{itemize}
    \item \textbf{Number of observations} identified per response.
    \item \textbf{Comparison} across different points in time or time intervals (e.g., between morning and afternoon).
    \item Whether the user reported \textbf{salient features}, such as peaks.
    \item Whether the user described points in time and time intervals \textbf{quantitatively} (e.g., ``8 a.m.'' or ``4p.m.~to 5 p.m.'').
    \item Whether the user reported time \textbf{qualitatively}, such as ``morning'' or ``late afternoon''.
\end{itemize}
\rev{Coders} assigned a numeric code for the number of observations and binary values for the remaining four codes to each response.
For the statistical analysis, we picked the average number of coded observations by the three independent coders (Fig.~\ref{fig:observations}).
For the other codes, we only picked 1 if at least two coders agreed, otherwise 0 (Fig.~\ref{fig:binary}).
From the qualitative time references, the coders extracted ten commonly mentioned day time periods (Fig.~\ref{fig:periods}). 
Each observation was then assigned to one or multiple time periods.

\begin{figure}[tb]
\centering
\includegraphics[width=0.8\columnwidth]{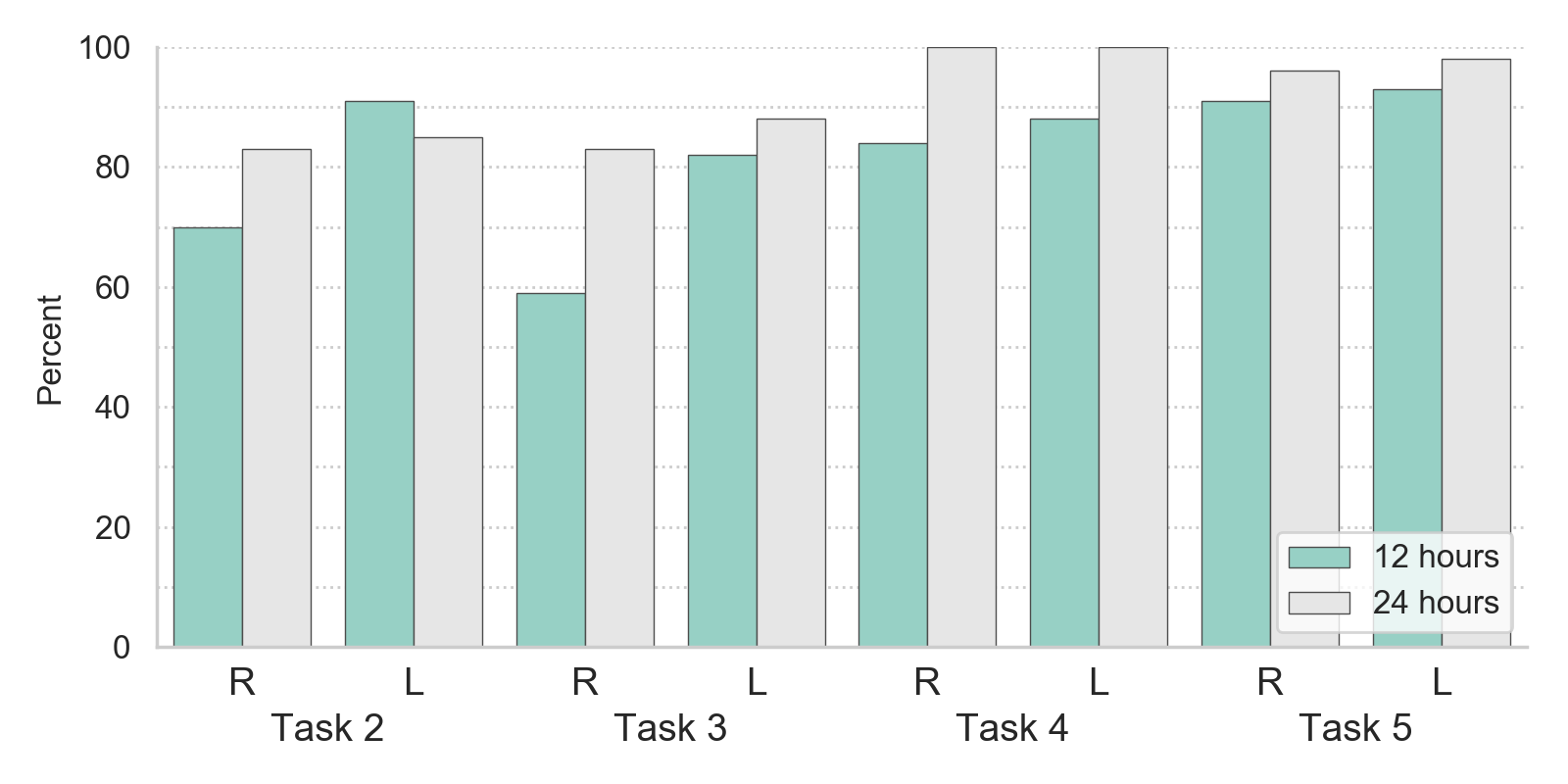}
\caption{Percentage of correct responses in tasks 2 --- 5 for radial (R) and linear (L).}
\label{fig:accuracy}
\end{figure}

The low-level analytic tasks (H2 -- H5) were evaluated with respect to accuracy by comparing the number of errors, and with respect to efficiency by comparing completion times.
Accuracy is visualized as the percentage of overall correct responses in Fig.~\ref{fig:accuracy}.
Error cases were removed for the completion time analysis and qualitatively explored to find potential explanations for the quantitative observations.
Results are plotted as means with 95\% confidence intervals \rev{in} Fig.~\ref{fig:completion_time_tasks}. 

Finally, \rev{we analyzed the} obtained \rev{subjective} ratings on a five-point Likert scale (H6) 
(Fig.~\ref{fig:task6_results}).
\rev{Each optional textual feedback was coded if it only contained negative, only positive, or both types of utterances ({\em i.e.}, neutral).}

\rev{All dependent variables were statistically analyzed using analysis of variance (ANOVA) with layout as within-subjects and cardinality as between-subjects factor.
Completion times were log-transformed.
For the analysis of binary codes in task 1, we used a Generalized Linear Mixed Model (GLMM) using a binary logistic regression model.}
All F-scores, descriptive statistics, details about the exploratory error case analyses, \rev{coded textual feedback, task descriptions, example stimuli etc.~}can be found in the supplemental \rev{study information document}. 

\section{Results}

In total, we had 92 users that were accepted for the HIT.
Each user performed each task once with linear, once with radial layout, so that we obtained 184 samples in total.
44 users performed cardinality 12, 48 users cardinality 24.

\subsection{Task 1: Untargeted Analysis}

\begin{figure}[h]
\centering
\includegraphics[width=0.6\columnwidth]{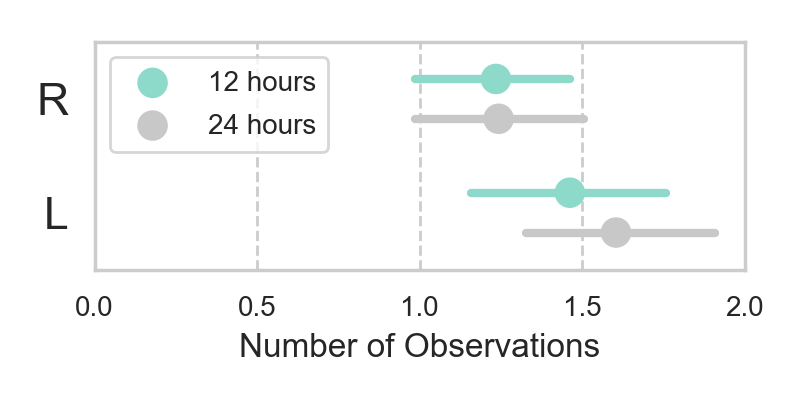}
\caption{Mean number of observations with 95\% confidence intervals in task 1 for radial (R) and linear (L).}
\label{fig:observations}
\end{figure}

\rev{An ANOVA} on the coded numbers of observations showed a \rev{medium} main effect for layout (\rev{$F_{1,90}=8.852;p=.004;\eta_p^2=.090$}) but not for cardinality (\rev{$F_{1,90}=.187;p=.667;\eta_p^2=.002$}), and no interaction between the factors (\rev{$F_{1,90}=.458;p=.500,\eta_p^2=.005$}).
On average, users reported 1.5 observations using the linear layout and 1.2 observations using the radial layout (Fig.~\ref{fig:observations}). 

From the 184 coded \rev{non-empty} responses, 24 (13\%) \rev{by 16 users} did not contain any observations about the data.
The responses without any data observations are almost equally distributed among the four conditions. 
From these 24 responses, 16 were contributed by eight participants, which indicates that they did not understand the task and reported irrelevant or superficial observations in both conditions.
The remaining eight were either task misinterpretations that only occurred in the first condition (three cases)\rev{, such as \emph{``The number of traffic accidents every hour?''} for 12l,} or explanations why the user did not come up with any data observations in the second condition (five cases), such as \emph{``the wheel was confusing, I didn't really understand any of it''} for 24r. 
\rev{As the 16 users did not perform worse than the other users in the remaining tasks, we did not exclude them from further analyses.}

\begin{figure}[h]
\centering
\includegraphics[width=0.8\columnwidth]{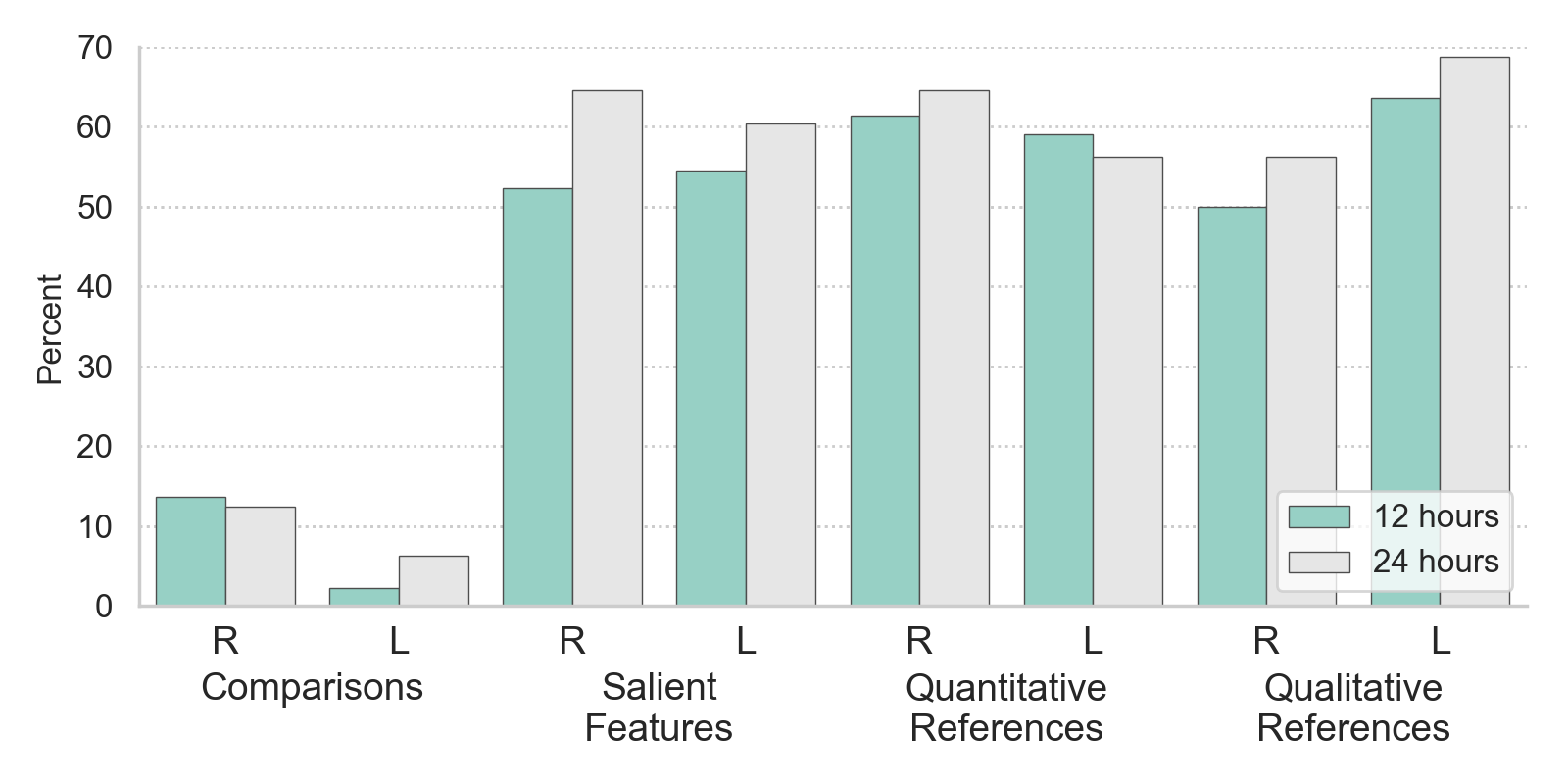}
\caption{Coded observations in percentage of all text responses in task 1 for radial (R) and linear (L).}
\label{fig:binary}
\end{figure}

The number of responses containing comparisons was generally low (16, which corresponds to 7\%).
There are more responses containing comparisons using the radial layout (12r: 14\%, 24r: 12.5\%) than using the linear one \rev{(12l: 2\%, 24l: 6\%)}. 
However, this difference \rev{did not reach statistical significance in our study}. 
Salient features were observed much more frequently (58\%), with a slight\rev{, non-significant} tendency towards 24-hours representations (60\% for 24l and 65\% for 24r) than 12-hours representations (54\% for 12l and 52\% for 12r).
The number of responses that contain quantitative time identifiers range between 56\% (24l) to 65\% (24r), and no factor \rev{was found to have} a significant influence on the usage of quantitative reporting of time.
However, we found a difference between qualitative time reporting for layout ($F_{1,180}=8.111;p=.005$): users reported times significantly more often qualitatively using a linear layout (12l: 64\%, 24l: 69\%) than using a radial layout (12r: 50\%, 24r: 56\%). 
The percentage of observations associated with the respective codes is visualized in Fig.~\ref{fig:binary}. 

\begin{figure}[h]
\centering
\includegraphics[width=0.8\columnwidth]{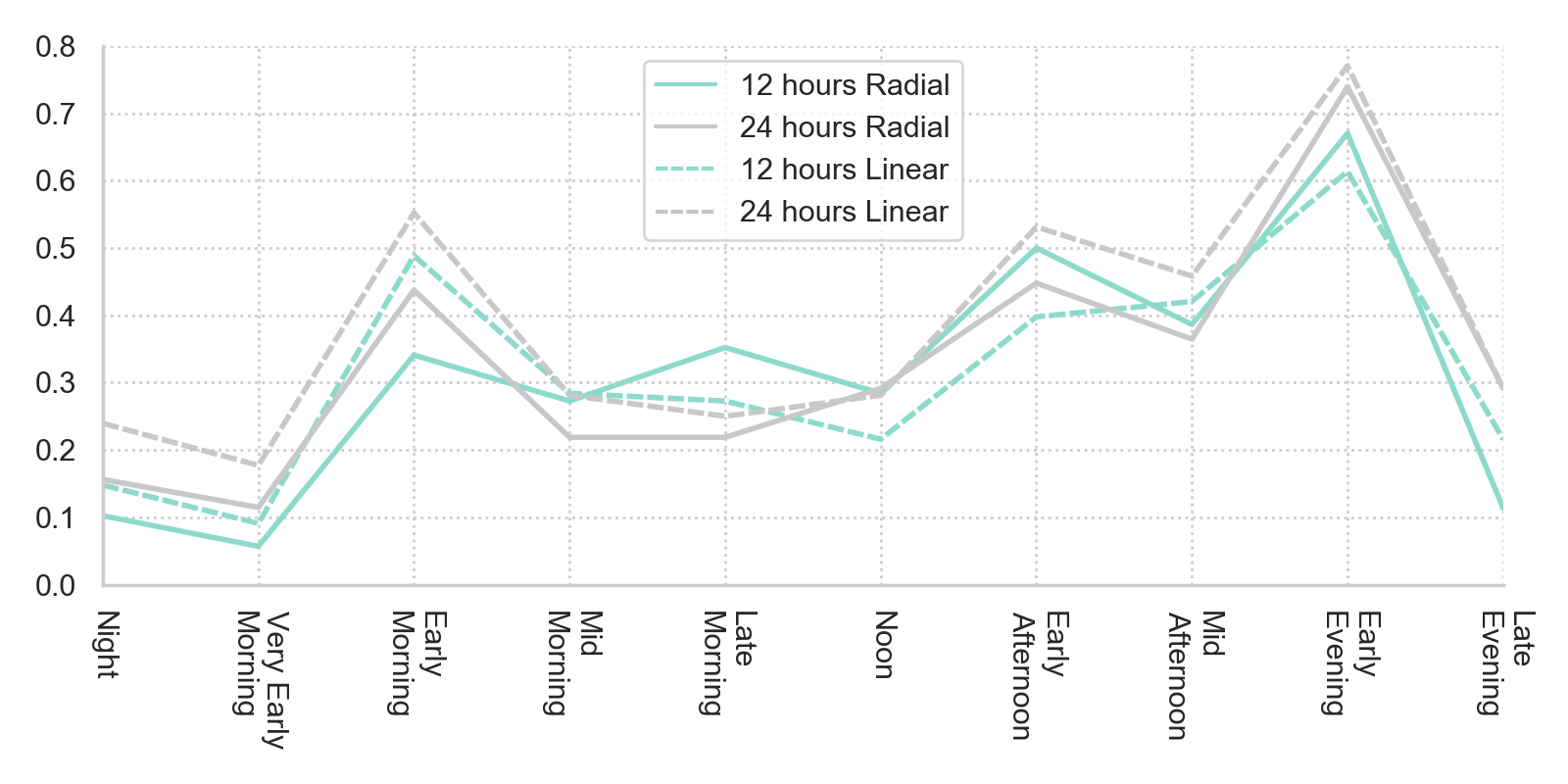}
\caption{Mean number of observations mentioning time periods of the day for the four conditions.}
\label{fig:periods}
\end{figure}

In Fig.~\ref{fig:periods}, we can furthermore observe some differences between the four conditions with respect to the time periods mentioned in the users' observations. 
First, it is clearly visible that the rush hour peaks were most commonly reported with the 24-hours variants, in particular 24l. 
Second, using 12l, users were reporting fewest observations about the afternoon rush hour. 
Third, using 12r, users reported few observations in the early morning, but more in the late morning than with any other visualization. 
Users did not report more about late evening or night hours using 24r, despite its continuous mapping across midnight. 

Hypothesis \textbf{H1} therefore is not supported: \emph{The 12-hours radial chart imitating a clock layout does not lead to a larger number of insights when viewing daily patterns for a few seconds. 
On the contrary, users reported slightly more observations and also used qualitative references to points in time (e.g., ``evening'') significantly more often with a linear layout than with a radial layout.} 

\begin{figure*}
\centering
\includegraphics[width=1.8\columnwidth]{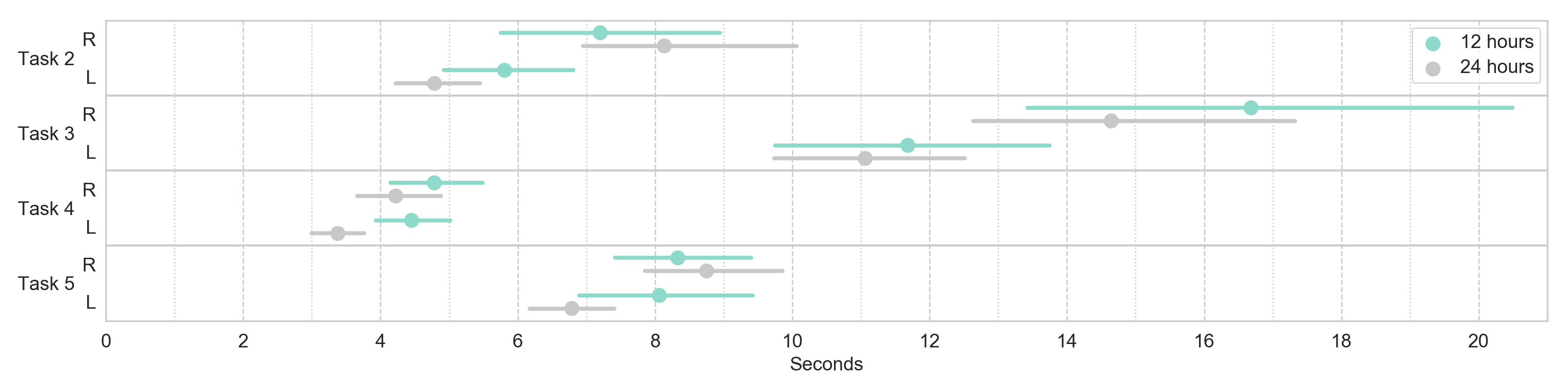}
\caption{Mean completion times with 95\% confidence intervals in task 2 (locate time), task 3 (read value), task 4 (locate maximum), and task 5 (compare a.m./p.m.~interval) for radial (R) and linear (L).} 
\label{fig:completion_time_tasks}
\end{figure*}

\subsection{Task 2: Locate Time}

We first computed the error by comparing the user's selected bar to the actual bar corresponding to the displayed label.
From the 184 total obtained samples 152 (82.6\%) were correct. 
The highest error rate was detected for 12r (30\%), and the lowest for 12l (9\%). 
Across all conditions, most of the errors (57\%) were caused by a 12 hours shift from the actual bar, {\em i.e.}, the users selected the corresponding a.m.~or p.m.~value or vice versa, while 23\% of responses are shifted by a single hour, and 10\% by a combination of switching a.m.~and p.m.~and shifting one hour ({\em i.e.}, 11 or 13 hours shift). 
Most of the single hour shifts arose using 24-hours variants (5 out of 7), and most a.m.~and p.m.~swaps occurred with 12r (8 out of 20). 
While these a.m./p.m. swaps by 12r were evenly distributed in both directions, all three a.m./p.m. swaps registered for 12l were incorrect selections from the a.m.~chart, {\em i.e.}, the upper chart, while the \rev{requested} time interval was given in p.m. 

For the completion time of correct responses, we found a \rev{large and} significant main effect for layout (\rev{$F_{1,61}=21.279;p<.001;\eta_p^2=.259$}) but not for cardinality (\rev{$F_{1,61}=.223;p=.638;\eta_p^2=.004$}). 
On average, users needed 7.4 seconds to select the correct bar with radial charts and 5.2 seconds for linear charts. 
There is some trend for 24-hours charts to be more effective than two 12-hours charts when using a linear layout (15\%, on average). 
There is a tendency for this trend to be reversed for radial charts, where the 24-hours variant was around 11\% less effective in terms of completion time than the 12-hours variant. 
This interaction is statistically significant \rev{with a small effect size ($F_{1,61}=4.504;p=.038;\eta_p^2=.069$)}. 
This means that our hypothesis \textbf{H2} is only partially supported: \emph{Juxtaposed 12-hours radial charts are slightly more efficient to locate a given time than a 24-hours radial chart, but they are more error-prone and less efficient than linear charts}.

\subsection{Task 3: Read Value}

\begin{figure}[tb]
\centering
\includegraphics[width=0.8\columnwidth]{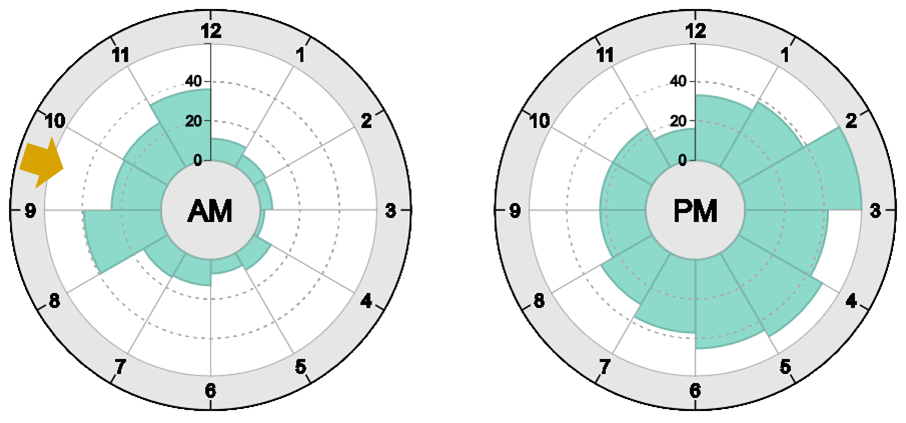}
\caption{Example for incorrectly read value with 12r (correct value is 25, user input was 7), which could be caused by an assumed grid line interval of five or by reading the value of the bar at 12-1 a.m. }
\label{fig:readValueErrors}
\end{figure}

From the total 184 samples, 
50 input values corresponded to the actual \rev{data} value. 
\rev{However, only 11 error cases differed by a value more than 4 from the actual data value, which is more than 10\% of our maximum data values.}
We therefore counted all input values as \rev{error} if they had an absolute deviation from the actual value of \rev{more} than 4 \rev{(11 samples) or were invalid ({\em i.e.}, empty or non-numeric, 29 samples)}. 
The lowest error was measured for 24l (14\%) and the highest for 12r (41\%!). 
From the 18 error cases of 12r, 12 were invalid inputs. 
When analyzing the valid error cases across all conditions, we could explain 45\% by rounding errors, where the user's input value lies within the same grid lines as the target value. 
The remaining error cases (four for 12r and one for 24r) can be potentially explained by one of the two following observations  (Fig.~\ref{fig:readValueErrors}): One error cause could be a misinterpretation of the visual encoding so that users reported the value of the closest bar to the axis label instead of the marked bar.
A second error cause may be the misinterpretation of grid lines, where a wrong assumption of the grid line intervals can cause incorrect value interpretations. 

To assess the efficiency of the charts, we performed \rev{an ANOVA} on completion time of \rev{the 64 users not committing any errors. We} found a \rev{large} main effect for layout (\rev{$F_{1,62}=21.064;p<.001;\eta_p^2=.254$}), but not for cardinality (\rev{$F_{1,62}=.220;p=.641;\eta_p^2=.004$}) and no interaction between the two factors (\rev{$F_{1,62}=.404;p=.527;\eta_p^2=.006$}). 
Overall, 12r required most time to come up with the correct response, and 24l led to the fastest response. 
This partially supports hypothesis \textbf{H3}: \emph{Reading values from a linear chart is less error-prone than reading values from a radial chart and also more efficient. 
However, there is no effective difference between 12-hours and 24-hours linear charts.}

\subsection{Task 4: Locate Maximum}

From all 184 cases, two responses were invalid and excluded from further analysis.
From the remaining 182 cases, there were 170 correct responses. 
For 24l and 24r, all recorded responses were correct.
For 12l, 88\% of the responses were correct, for 12r 84\%. 
We qualitatively analyzed all 12 error cases and classified the potential reasons for the errors into three categories.
The most common reason (five for 12l and four for 12r) was that users selected the maximum value from the incorrect chart -- {\em i.e.}, from a.m.~instead of p.m.
For the three remaining error cases of 12r, users selected either the second highest bar (two cases) or the highest value next to one of the value axes (one case). 

The \rev{ANOVA} on completion time showed \rev{small but} significant main effect\rev{s} for layout (\rev{$F_{1,79}=7.507;p=.008;\eta_p^2=.087$}) and for cardinality (\rev{$F_{1,79}=5.721;p=.019;\eta_p^2=.068$}), but no interaction (\rev{$F_{1,79}=2.742;p=.102;\eta_p^2=.034$}). 
On average, the correct maximum was selected with a linear layout in 3.9 seconds and with a radial layout in 4.5 seconds.
Cardinality 24 led to significantly faster selection times (3.8 seconds) than cardinality 12 (4.6 seconds). 
Overall, the maximum was selected fastest using 24l (3.4 seconds). 
Hypothesis \textbf{H4} is therefore supported:
\emph{Identifying the maximum value is most efficient using a single, linear 24-hours representation of the time series data.}

\subsection{Task 5: Compare AM/PM~Interval Values}

We counted every incorrect response as error, which are 10 cases in total.
We found the highest number of incorrect responses for 12r (9\%) and the lowest for 24l (2\%). 
In four cases, users reported that the indicated values are equal when their value differences were rather low (1-7). 
\rev{This means, users underestimated the value differences.}
These error cases only happened with 12r (2) and 12l (2). 

\rev{The ANOVA} on completion time showed a \rev{medium-sized} main effect for layout (\rev{$F_{1,80}=11.516;p=.001;\eta_p^2=.126$}), but not for cardinality (\rev{$F_{1,80}=.216;p=.644;\eta_p^2=.003$}). 
We also found a \rev{medium-sized} interaction between the two factors (\rev{$F_{1,80}=8.393;p=.005;\eta_p^2=.095$}). 
On average, the correct comparison was performed within 7.3 seconds for linear layouts and 8.6 seconds for radial layouts.
The fastest response was given with 24l (6.8 seconds), the slowest with 24r (8.8 seconds). 
\rev{The completion time of the two 12-hours variants was very similar (8.1 and 8.3 seconds).} 
Therefore, hypothesis \textbf{H5} is partially supported: \emph{
\rev{Linear 24-hours charts facilitate the comparison of a.m./p.m.~interval values.}
The 12-hours linear chart is not superior to the radial charts. }

\subsection{Task 6: Subjective Rating}

\begin{figure}[h!]
\centering
\includegraphics[width=0.6\columnwidth]{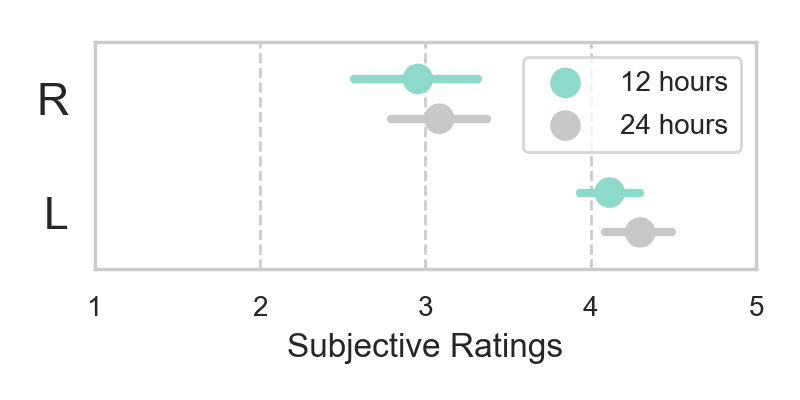}
\caption{Mean ratings issued on a five-point Likert scale with 95\% confidence intervals.}
\label{fig:task6_results}
\end{figure}

For the ratings, we found a \rev{large} main effect for layout (\rev{$F_{1,88}=83.912;p<.001;\eta_p^2=.488$}), but not for cardinality (\rev{$F_{1,88}=.548;p=.461;\eta_p^2=.006$}), and no interaction (\rev{$F_{1,88}=.156;p=.694;\eta_p^2=.002$}). 
On average, the linear charts received \rev{ratings} 1.2 points higher than the radial charts on the 5-point Likert scale (see Fig.~\ref{fig:task6_results}).
24l received the highest average \rev{rating} (4.3) and 12r lowest (2.95). 

To better understand the ratings, we analyzed the users' optional text comments. 
We received 66 text comments in total -- most for 24r (21, which is a 43\% comment rate) and fewest for 12l (27\% comment rate). 
24r received most \rev{purely} negative comments (\rev{66\%}), followed by 12r (\rev{50\%}). 
The 24-hours radial chart was mainly criticized for its unusual appearance with comments like:
\emph{``It's actually a useful way to show the info, but is unusual and could be confusing at first.''}
or:
\emph{``It is not easy to interpret. I know it follows a clock format but it doesn't translate well.''} 
The 12-hours radial charts received similar comments, like:
\emph{``This is really hard to look at until you realize that it is showing like a clock.
But the data is still difficult to decipher.''}

12l received more \rev{purely} positive comments (\rev{50\%}).
24l \rev{did not receive any purely negative comments}. 
Compared to the juxtaposed 12-hours radial charts, the linear bar charts were 
easier to read, e.g.:
\emph{``Now that I've seen both, this method is easier and more intuitive than the clock type visualization.
It's easier to look at the bar charts and digest the information.''}
The 24-hours bar chart was primarily appreciated for its effectiveness and simplicity, for instance because \emph{``it seems clear and simple, very easy to comprehend at a glance.''}
\rev{A list of all user comments can be found in the supplemental material.}
This only partially confirms hypothesis \textbf{H6}:
\emph{Both radial charts (12r and 24r) received significantly lower subjective ratings and more negative comments than the linear charts}.

\section{Discussion}

Our main hypothesis that daily temporal patterns can be interpreted more effectively using clock-like 12-hours rose charts is not supported.
Overall, the conventional and most elementary continuous linear bar chart performed more accurately and efficiently across all low-level tasks.
This is particularly remarkable as the linear 24-hours bar chart requires least ink and space to encode the data.
We discuss the results with respect to the experimental factors and our main hypothesis. 


\subsection{Linear vs.~Radial Layout}

\rev{Summarized, the independent variable} layout can explain \rev{most} observed performance differences: 

\begin{itemize}
    \item Using linear charts, users reported more observations about the data, and these observations more often used qualitative time references than observations derived from radial data visualizations. 
    \item All low-level tasks could be solved significantly faster using a linear layout than \rev{both radial layouts, including finding a specified hour of a day}. 
    \item Participants clearly preferred linear charts. Many users reported that radial charts are harder to read, and some even reported fundamental difficulties interpreting the visual encoding. 
\end{itemize}

To a certain extent, these results were expected \rev{and generalize findings} that reading values, locating the maximum and comparing values is \rev{less} accurate and efficient using \rev{radial bar and line charts}~\cite{goldberg2011eye,brehmer2017visualizing} \rev{to rose charts.}
After analyzing error cases and user feedback, we speculate that \rev{rose} charts are sometimes decoded fundamentally wrong, for instance by not knowing how the indicated bars' length corresponds to the axis on the top or by assuming a wrong grid line interval. For instance, one user reported that \emph{``I JUST noticed that each bar represents 20 accidents. Before I thought 1 bar = 1 accident.''}
This can also explain the slightly lower \rev{performance} 
in task 1, as some users probably had problems interpreting the visualized data without guidance how to read the chart. 
Radial charts are not uncommon.
Yet, it seems that conventional linear bar charts are much more ubiquitous than radial charts\rev{ -- even more than classic pie charts \cite{spence2005no}.} \rev{Consequently,} they \rev{probably} can be \rev{understood} by a broad audience without any instructions. 
This is supported by various users' feedback, such as 
\emph{``this is a form that's more familiar -- no time really needed to understand the format of the data''} [24l]. 

On the other hand, some users also commented that they found radial charts more appealing or interesting with comments like \emph{``this} [24l] \emph{is simpler, less pretty or fancy but more effective than the previous one''} [24r] or 
\emph{``\rev{[...]} I just think the other one} [24r] \emph{is much more fun.''}
This supports Borkin~et~al.~\cite{borkin2013makes} who speculate that traditional linear charts are less engaging, which could explain why they are least memorable for a non-expert audience. For hierarchical data, Mer{\v{c}}un~\cite{mervcun2014evaluation} also showed that linear, intended lists are perceived as more useful and efficient, but less engaging and appealing than radial sunburst and circle pack diagrams. 
\rev{These findings have} a high relevance for the field of infographics, where designers seek to create aesthetically pleasing visual representations \rev{and less often use conventional bar or line charts \cite{borkin2013makes}}. 
Our results reinforce that designers of visualizations for the masses should stick to very simple and ubiquitous visual encodings to lower the risk of data misinterpretation due to incorrect decoding.

\subsection{2$\times$12-Hours vs.~1$\times$24-Hours Chart}

The separation into two less cluttered charts had much less influence on the completion time results than the layout. However, it has some negative influence on the data interpretation: 

\begin{itemize}
    \item Locating the maximum is most efficient using a continuous, linear chart and also more error-prone using \rev{separate} charts. This is also indicated by a slightly higher number of salient features reported using 24-hours variants in the untargeted analysis. 
    \item There is a small tendency to underestimate value differences between bars in two separate charts. 
\end{itemize}

Even though the linear 24-hours bar chart was the only visualization that did not benefit from the fact that we compared associated a.m./p.m.~intervals, it performed better than the radial charts and the separate 12-hours linear bar charts \rev{in task 5}.
\rev{For both, linear and radial} separate charts, we found that users \rev{underestimated value differences.}
This reinforces that lengths comparisons are easier when the bars are horizontally aligned, thereby confirming previous results~\rev{\cite{cleveland1984graphical,simkin1987information,heer2010crowdsourcing,talbot2014four}}.

It seems that users direct\rev{ed} more attention to the a.m.~data using the linear 12-hours chart than with any other visualization. 
\rev{This is indicated by the following observations with 12l: 1) users reported relatively more observations in the morning than in the afternoon in task 1 (Fig.~\ref{fig:periods}), 2) some users mistakenly selected a.m.~intervals instead of p.m.~intervals in task 2, but never vice versa, and 3) users sometimes selected the maximum value of the a.m.~chart instead of the p.m.~chart in task 4, but never vice versa.
Diehl et al.~\cite{diehl2010uncovering} report a reading direction effect from top-left to bottom-right in Cartesian charts, which could be an explanation for these findings. 
We therefore assume that the performance difference between the 12- and 24-hours variants of linear charts was caused by the separation of a.m.~and p.m.~values rather than the different bar widths.}
Separating a \rev{bar} chart into meaningful sub-units therefore can be a way to direct the user's attention to the data encoded in the top chart. However, if accurate value comparisons or understanding of global maxima are required, a continuous, linear chart is the best option.

\subsection{Clock Layout}

Finally, we discuss the results concerning the hypothesis that clock-like charts facilitate the cognition of daily patterns: 
\begin{itemize}
    \item Across all low-level tasks, users committed most errors using the two juxtaposed 12-hours radial charts. 
    \item Locating one-hour day-time intervals is slightly more efficient using two 12-hours radial charts than one 24-hours radial chart, but more error-prone and less efficient than if using linear charts. 
    \item The 24-hours radial chart received more negative subjective feedback than the 12-hours radial chart. 
\end{itemize}



One particularly unexpected finding was that users were significantly more efficient locating a given one-hour time interval using a linear layout than using a radial layout. This contradicts results by Fuchs~et~al.~\cite{fuchs2013evaluation}, which indicate that a radial layout is more suitable for detecting day-time locations. Partially, the poor performance of radial charts can be explained by the high completion times for the 24-hours radial chart (Fig.~\ref{fig:completion_time_tasks}, first row). 
This was expected, since the 24-hours radial chart does not use the common clock layout. 
\rev{A major difference to the study by Fuchs et~al.~\cite{fuchs2013evaluation} is that our users were presented with large-scale visualizations instead of a matrix of glyphs. Therefore, our visualizations had labels indicating the precise hour of the day, which can be an explanation why users could locate the specified time efficiently in the linear charts compared to the unlabeled linear stripe and line charts by Fuchs et~al.} 

\rev{Most of the errors when locating the time using the radial 12-hours variant was} 
caused by swapping a.m.~and p.m. This error also occurred with the linear 12-hours chart, but less frequently, and only if the time interval to be selected was p.m. 
In contrast to the two linear bar charts, two juxtaposed rose charts do not support the natural reading order. In the original rose charts of the mortality in the Crimean War, Florence Nightingale connected March and April 1855 in her two juxtaposed charts with a line. In our study, the two 12-hours rose charts were implicitly connected by noon \emph{and} midnight, which are located on the top. This ambiguous meaning of 12 on the top was perceived as difficult to understand: 
\emph{``I found it very hard to read initially as I was trying to figure out if the 12pm on the left was for PM or AM, based on it stating AM in the center. This is very confusing.''} 
We also observed an increased tendency to report time periods just before noon in task 1 using the separate 12-hours charts compared to the other visualizations (Fig.~\ref{fig:periods}). This could indicate that the users' attention is strongly focused to the point in time where the chart is separated -- even though there is no salient data point.
\rev{We speculate that this effect stems from the artificial discontinuity between the non-neighboring data points rendered next to each other (e.g., Fig.~\ref{fig:12r}). 
To systematically test this hypothesis, an eye tracking study comparing separate rose charts with changing variability between neighboring bins will be an interesting future work.}

The separation of the 24 data bins into two juxtaposed charts also affects the appearance of the bars. The segments of the single 24-hours variant were narrower and the blue areas therefore looked more like bars than circle segments. Maybe it was perceived -- and therefore also read -- like a traditional bar chart by more users. 

From the user feedback, we suspect that some users also did not draw the connection between a clock layout and the 12-hours rose charts -- at least not initially.
But even users who explicitly stated that they did see the connection between a clock and the rose chart reported that they found it hard to read.
Rose charts for daily patterns use the same layout but a different encoding than analog clocks. 
It could be that people are trained to read the orientation of the clock hand rather than to memorize the circle segment corresponding to a particular hour. 
This way, they cannot translate their clock reading skills to interpret daily patterns or even to locate a given hour on a ring separated in twelve segments. 
Reading the clock itself is also considered a cognitively demanding task, and a common task to assess cognitive impairments is to ask patients to draw clocks showing a given time~\cite{royall1998clox}. 

Only few users explicitly appreciated the clock layout, for instance \emph{``I like this visualization because I can easily understand it based on intuition. It looks like a clock.''} or \emph{``I think reading particular numbers of this graph} [12l] \emph{is easier, but the clock graph} [12r] \emph{gives a better feel for time of day.''} 
However, some users also associated the 24-hours radial chart with a clock: \emph{``It's neat in that it parallels an analog clock face.''} 
It therefore seems that it is mainly the radial layout that creates the impression of reading a clock, not necessarily the radial arrangement of 12 hours. 

In summary, 12-hours radial charts combine the disadvantages of radial charts (inaccurate value reading, inefficient location of global maxima, inefficient value comparisons, and unfamiliarity) and separate charts (higher risk of reading the wrong chart and tendency to underestimate value differences). In addition, the meaning of the top 12 is ambiguous, and the bars get more distorted due to the larger circle segments compared to the 24-hours variant. The connection to the clock layout was also not apparent to many users and even if so, they reported problems utilizing the clock reading skill to interpret rose charts. The assumed intuitiveness of visualizing daily patterns in an analog clock layout could not be shown in this study.  

\section{Study Limitations}

A validity threat of every crowd-sourced study is the lack of control over the \textbf{environment}.
One potential confounding factor in our study is the browser window size of the users, which affects the four conditions differently, depending on the size of the visualization.
We did not query the window size, but even the largest visualization ({\em i.e.}, 12r) is fully visible on a common mobile phone screen in portrait mode. 
Also, we have very few error cases in task 5, where both charts had to be interpreted. 
We can assume that both charts were visible for our users.

We selected one radial \textbf{encoding} to visualize daily patterns which is inspired by very early work by Florence Nightingale~\cite{cohen1984florence, brasseur2005florence}. However, there are other encodings that can be used with a linear and radial layout (Table~\ref{tab:designSpace}). 
Using color, for instance, there is also the possibility to use a spiral display~\cite{dragicevic2002spiraclock,tominski2008enhanced}, which is also interesting as it manages to overcome the problem of two juxtaposed charts. 

As our goal was to simulate charts as found in news or public websites, we used a fixed, representative \textbf{size} for all visualizations. However, temporal patterns are sometimes also presented as glyphs \cite{fuchs2013evaluation} or sparklines \cite{beck2017word}. As these small-scale visualizations usually only convey a general trend, the low-level analysis tasks used in our experiment are not appropriate for these visualizations. 

We used only one type of \textbf{data} in our study. Based on prior results showing that different data sets do not interact with the layout \cite{brehmer2017visualizing}, we assume that data with the same characteristics (e.g., daily network traffic~\cite{erbacher2002intrusion,kintzel2011monitoring}, mobile phone usage~\cite{shen2008mobivis}, or traffic information~\cite{huang2016trajgraph}) would lead to comparable results.
However, previous work suggests an interaction between layout and bin size for different time periods~\cite{fuchs2013evaluation,brehmer2017visualizing}, where increasing the data granularity and the number of bins, respectively, improves the relative performance with radial charts. Thus, \rev{it would be interesting to investigate} daily patterns \rev{showing} a finer granularity, such as 15-minutes bins, \rev{for radial charts in the future}. 


Finally, we performed our experiment with non-expert \textbf{users}. 
Subjective feedback indicates that participants got more used to radial charts over the short duration of the experiment.
\rev{However, the self-reported visualization literacy of our participants did not have an influence on the number of reported observations or subjective ratings.
It is therefore unclear} whether rose charts are suitable for visualizing daily patterns for trained expert users. 
\rev{Also, some users may not be confronted with analog clocks so frequently and may require more training to get familiar with the clock layout.}


\section{Conclusions}

\rev{Complementing prior work on radial charts for visualizing monthly and weekly periodical data \cite{fuchs2013evaluation, brehmer2017visualizing}, we showed that radial charts also do not improve the understandability of \emph{daily} periodical data -- despite their resemblance with an analog clock. Our study also extends previous results, which showed that radially arranged bars can be decoded less accurately and efficiently than linearly arranged bars in many low-level analytical tasks \cite{goldberg2011eye,brehmer2017visualizing}, to rose charts. Furthermore, we showed that chart separation makes the detection of salient features slightly more difficult, and may lead to underestimation of value differences.}
Based on these observations, we provide two recommendations for designers.

    First, visualizations for the masses should rely on linear bar charts instead of radial rose charts even for showing periodical daily data. \rev{Shaping a rose chart like a clock does not improve the users' understanding of daily patterns.}

    Second, continuous bar charts should be preferred over separate ones, especially when accurate data interpretation is important. However, when horizontal space is limited, it is more advisable to separate a linear bar chart into multiple charts rather than to use a single rose chart. \rev{Separation of radial charts should be avoided, as it can cause unwanted attention effects.}
In summary, we have contributed further evidence against the usage of radial charts, in particular rose charts.
However, there might be other data and application domain cases, such as the analysis of wind directions, where rose charts can be highly beneficial. 




\acknowledgments{
This work was partially financed by the Austrian Science Fund (FWF): T 752-N30 and by grant PGI 24/K061 of the Universidad Nacional del Sur (Argentina).
This work was partially written in scope of the COMET program at VRVis (854174).
The authors would like to thank Gemza Ademaj for open coding support.}

\balance

\bibliographystyle{abbrv-doi}
\bibliography{tcv}
\end{document}